\documentclass[epj]{svjour}
\usepackage{graphics}
\begin{document}
\title{Strange dibaryon and $\bar{K}NN$-$\pi \Sigma N$ coupled channel equation}
\author{Y. Ikeda \and T. Sato
%\author{Y. Ikeda\inst{1} \and T. Sato\inst{1}% etc
}                     % Do not remove
\institute{Department of Physics, Graduate School of Science,
Osaka University, Toyonaka, Osaka 560-0043, Japan}
\date{Received: date / Revised version: date}
\abstract{
$\bar{K}NN$ three body resonance has been studied by 
$\bar{K}NN-\pi\Sigma N $ coupled channel Faddeev equation.
The S-matrix pole has been investigated 
using the analytically continued scattering amplitude
on the unphysical Riemann sheet.
As a result we found a three-body resonance of 
strange dibaryon system with the
binding energy and width $B \sim 76$MeV and $\Gamma \sim 54$MeV.
 % end of PACS codes
} %end of abstract
\maketitle
\section{Introduction}
\label{intro}

The meson-nucleus bound state  has been an important
tool to study the meson properties inside nuclear medium and
the interaction of meson below threshold energy.
The $\bar{K}$-nuclear  system is particularly interesting because of the
 $I=0$ resonance $\Lambda(1405)$ below $\bar{K}N$ threshold.
The attractive kaon-nucleus interaction obtained 
from the the analysis of the X-ray from the kanonic-atom\cite{Gal1,Ram} 
might be largely related to the $\Lambda(1405)$.
In a few nucleon system, where one hopes to learn
about the kaon-nucleon interaction with less ambiguity on nuclear 
many body dynamics, possible deeply bound states of
the kaon in nuclei has been proposed by Akaishi and 
Yamazaki\cite{Aka1,Aka2,Dote}. The kaon-nucleus optical potential
is constructed from the kaon-nucleon g-matrix in nuclear medium.
The predicted binding energy $B$ and width $\Gamma$ of the
smallest nuclear system $K^-pp$ is $(B,\Gamma) = (48,61)$MeV.
FINUDA collaboration reported a signal of the
$K^-pp$ bound state from the analysis of the invariant mass distribution
of $\Lambda-p$ in the $K^-$ absorption reaction on nuclei\cite{Agne}.
The reported central value of the binding energy
is  $(B,\Gamma)=(115,67)$MeV, which 
is twice larger than the theoretical  prediction.
Recently a  question raised\cite{Oset2} on the
$\bar{K}N-\pi\Sigma$ interaction used in  Ref. \cite{Aka1,Aka2}, where
the interaction in $\pi\Sigma$ channel is absent, and 
on the  interpretation of the invariant mass spectrum\cite{Magas},
which might be explained by the $K^- p p \rightarrow \Lambda p$ reaction 
with final state interaction.

%-------------------------------------------------------
Intuitively, the $K^-pp$ resonance may be regarded as
the bound state of $\Lambda(1405)$ and nucleon 
interacting with kaon exchange mechanism.
The binding energy of this resonance will be strongly influenced  
by the dynamics of the $\Lambda(1405)$ resonance. 
For the resonance interaction in a few body system,
 it will be very important to take into account
fully the kaon-nucleon dynamics in the $\bar{K}NN$-three body system
including the decay of the $\Lambda(1405)$ into the $\pi\Sigma$ state.
The purpose of this work is to study the
 strange dibaryon system by taking into account the three-body
dynamics using the $\bar{K}NN-\pi\Sigma N$ coupled channel Faddeev equation.
The  three-body resonance
have been investigated on the three-neutron\cite{Glock,Moll},
 $\pi NN$ dibaryon\cite{Matsu} and $\Sigma NN$ hypernuclei\cite{Pear,Afnan}.
The resonance can be studied  from 
 the pole of the S-matrix or scattering amplitude.
The pole position can be obtained by studying the
eigenvalue of the kernel of scattering equation,
which is analytically continued in the unphysical sheet.
We briefly explain our $\bar{K}NN-\pi\Sigma N$ coupled channel equation 
and the procedure to search the three-body resonance in section 2.

%-----------------------------------------------------------
The structure of $\Lambda(1405)$ have been a long standing issue.
The chiral Lagrangian\cite{Jido,Bor} approach is able 
describe well  the low energy $\bar{K}N$ reaction.
A genuine $q^3$ picture of  the $\Lambda(1405)$ 
 coupled with meson-baryon\cite{Hama} 
may  not be yet  excluded.
In this work we describe a $\bar{K}N-\pi\Sigma$ state
using  s-wave meson-baryon potentials
guided from the lowest order chiral Lagrangian.
With this procedure, the strength of the potentials
and the relative strength of the potentials among various 
meson-baryon channels are not parameters but determined
from the chiral Lagrangian. In this model, 
$\Lambda(1405)$ is 'unstable bound state', whose
pole in the unphysical sheet will become bound state of $\bar{K}N$
when the coupling between $\bar{K}N$ and $\pi\Sigma$ is turned off.
The model of the two-body meson-baryon 
interaction used in this work is explained in section 3.
We then report our results on the $\bar{K}NN$ dibaryon resonance in section 4.

\section{Coupled channel Faddeev equation and resonance pole}

In this section we briefly explain our coupled channel
equation and a  method to find resonance
pole from the coupled channel Faddeev equation.
Our starting point is the  Alt-Grassberger-Sandhas equation\cite{ags} for the
three-body scattering problem.
The AGS equation for three body scattering amplitude
$U_{ij}$  is given as
\begin{eqnarray}
U_{i,j} & = & (1 - \delta_{i,j})G_0^{-1} +
\sum _{n\neq i}t_n G_0U_{n,j} ,\  \label{Eqags-1}
\end{eqnarray}
Here we represent the spectator particle $i=1,2,3$ and the 
interacting particles $j,k$.
Scattering t-matrix for particle $j,k$ is denoted as $t_i$ and
$G_0 = 1/(W - H_0 + i\epsilon)$ is three particle Greens function.
 
With  the separable two-body interaction given as
\begin{eqnarray}
v_i & = & |g_i>\gamma_i < g_i|,
\end{eqnarray}
the AGS-equation in Eq. (\ref{Eqags-1}) takes the following form
\begin{eqnarray}
X_{i,j} & = & (1-\delta_{i,j})Z_{i,j} + \sum _{n\neq i} \int d \vec{p}_n 
    Z_{i,n}\tau _n X_{n,j}. \label{Eqags-2}
\end{eqnarray}
The amplitude $X_{ij}$ is  the matrix element of 
$U_{ij}$ between states $G_0|\vec{p}_i,g_i>$
with the plane wave spectator $|\vec{p}_i>$ and 
and  the interacting pair $|g_i>$, as
\begin{eqnarray}
X_{i,j} & = & <\vec{p}_i,g_i|G_0U_{i,j}G_0|\vec{p}_j,g_j>.
\end{eqnarray}

The driving term $Z_{ij}$ of Eq. (\ref{Eqags-2})
is particle exchange interaction defined as
\begin{eqnarray}
Z_{i,j} & = & <\vec{p}_i,g_i|G_0|\vec{p}_j,g_j>.
\end{eqnarray}
The 'isobar' propagator $\tau_i$ is given as
\begin{eqnarray}
\tau(W) & = & [1/\gamma_i - \int d\vec{q}_i 
\frac{<g_i|\vec{q}_i><\vec{q}_i|g_i>}
{W - E_i(\vec{p}_i) - E_{jk}(\vec{p}_i,\vec{q}_i)}]^{-1},
\end{eqnarray}
where $E_i$ and $E_{jk}$ are energy of the spectator and
the interacting pair, respectively.  $\vec{q}_i$ is relative
momentum of the pair $j,k$.

In our  $\bar{K}NN$ resonance problem, we have included
following $\bar{K}NN$ and $\pi\Sigma N$ Fock space components,
\begin{eqnarray}
| a > & = &  | N_1, N_2, \bar{K}_3>,\\
| b > & = &  | N_1, \Sigma_2, \pi_3>,\\
| c > & = &  | \Sigma_1, N_2, \pi_3> .
\end{eqnarray} 
After symmetrizing the amplitude using the $N_1$ and $N_2$ are
the identical particles\cite{Afnan2} and the partial wave
expansion of the amplitudes\cite{book} restricting
s-wave, the AGS-equation reduces into following
coupled integral equation,
\begin{eqnarray}
X_{l,m}(p_l,p_m) & = & Z_{l,m}(p_l,p_m) +\sum_n \int dp_n p_n^2
\nonumber \\ &\times&
 K_{l,n}(p_l,p_n) X_{n,m}(p_n,p_m). \label{Eqags-3}
\end{eqnarray}
Here we used simplified notation for kernel $K=Z\tau$.

We follow the method for searching the three-body resonance
used by Matsuyama and Yazaki\cite{Bal,Moll,Matsu}.
The AGS-equation of Eq. (\ref{Eqags-3}) is 
 the Fredholm type integral equation with the kernel $K=Z\tau$.
Using the eigenvalue $\eta_a(W)$ and eigenfunction $|\phi_a(W)>$
 of the kernel for given energy $W$,
\begin{eqnarray}
Z\tau|\phi_a(W)> & = & \eta_a(W)|\phi_a(W)>.
\end{eqnarray}
the scattering amplitude $X$ can be written as
\begin{eqnarray}
X & = & \sum_a \frac{|\phi_a(W)><\phi_a(W)|Z}{1 - \eta_a(W)}.
\end{eqnarray}
At the energy $W=W_p$ where $\eta_a(W_p)=1$, the amplitude
has a pole and therefore $W_p$ gives 
the bound state or resonance energy.

\section{Model of the meson-baryon interaction}

We investigate the strange $S=-1$ dibaryon state with
the total angular momentum $J=0$, parity $\pi=-1$ and iso-spin  $I=1/2$,
which is expected to have larger $I=0$ $\bar{K}N$ component than spin
triplet state.  
The s-wave meson-baryon ($\bar{K}N-\pi\Sigma$, $\pi N$)
interactions and baryon-baryon interactions are included.
Here we explain  on our model of the 
most important $\bar{K}N$ interaction.

The leading order chiral effective Lagrangian
for the octet baryon $B$ and the  pseudoscalar meson $\psi$ fields is
given as
\begin{eqnarray}
L_{int} & = & \frac{i}{8F_\pi^2}tr(\bar{\psi}_B \gamma^\mu
                                 [[\phi,\partial_\mu \phi],\psi_B]).
\end{eqnarray}
The meson-baryon potential derived from the chiral Lagrangian can be written as
\begin{eqnarray}
<\vec{p}',\alpha|V_{BM}|\vec{p},\beta>
 & = & - C_{\alpha,\beta}\frac{1}{(2\pi)^3 8F_\pi^2}
\frac{E_{M'}(\vec{p}')+ E_M(\vec{p})}
{\sqrt{4E_{M'}(\vec{p}')E_M(\vec{p})}} \nonumber \\ & \times &
v_\alpha(\vec{p}')v_\beta(\vec{p}).
\end{eqnarray}
Here $\vec{p}'$ and $\vec{p}$ are the momentum of the meson in the
initial state $\beta$ and the final state $\alpha$.
The strength of the potential at zero momentum is determined
by the pion decay constant $F_\pi$ and the relative strength among
the meson-baryon states is given by the constants $C_{\alpha,\beta}$,
which are $C_{\bar{K}N-\bar{K}N}=6,C_{\bar{K}N-\pi\Sigma}=\sqrt{6}$
and $C_{\pi\Sigma-\pi\Sigma}=8$.
The only parameter of our model is cut off $\Lambda$
of the phenomenologically introduced vertex function 
$v_\alpha(\vec{p})=\Lambda_i^4/(\vec{p}^2 + \Lambda_i^2)^2$.

\begin{figure*}
\resizebox{1.0\textwidth}{!}{%
\includegraphics{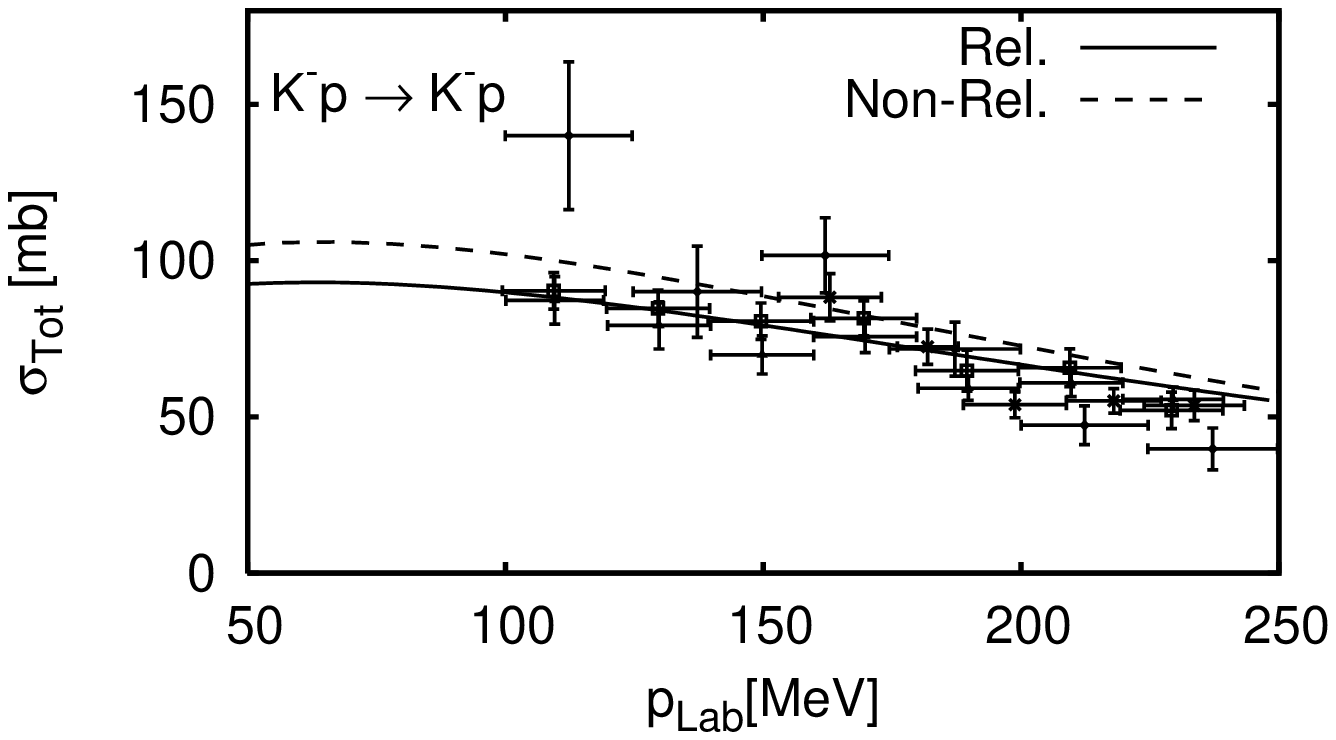}\hspace*{0.5cm}
\includegraphics{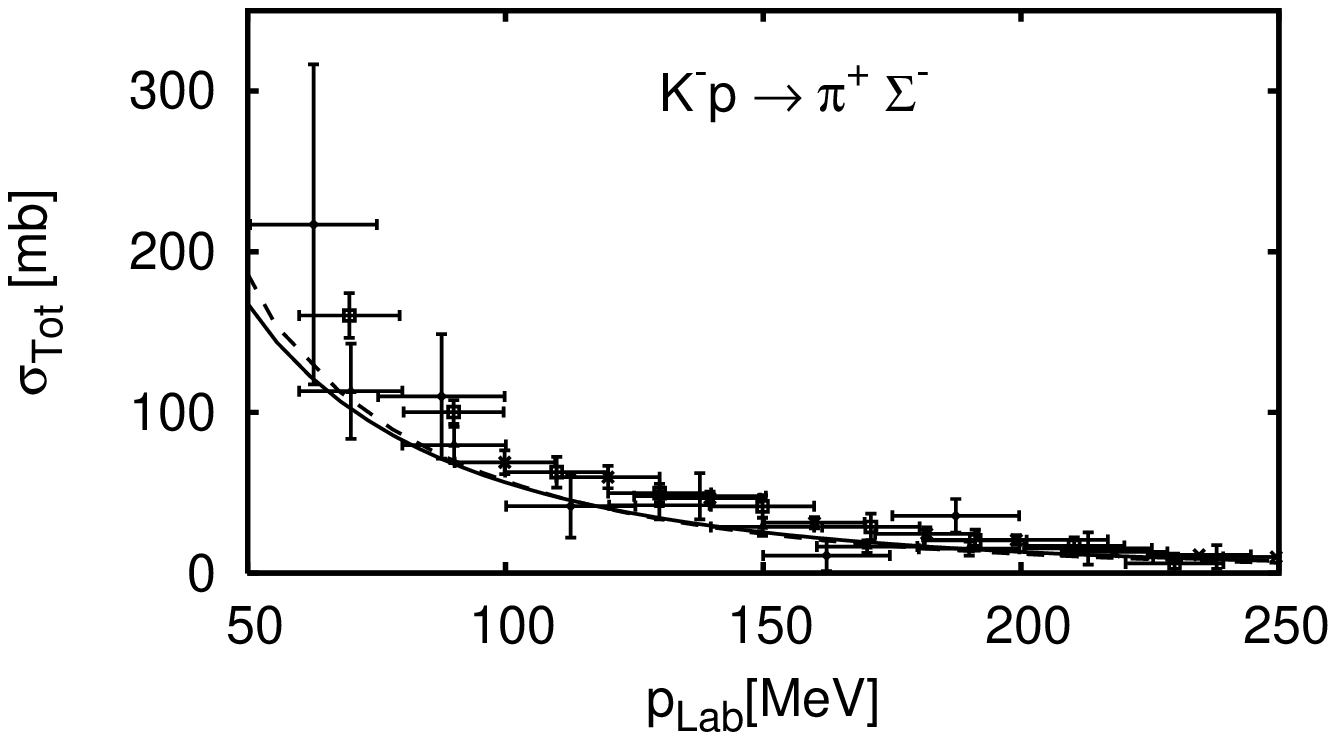}\hspace*{0.5cm}
\includegraphics{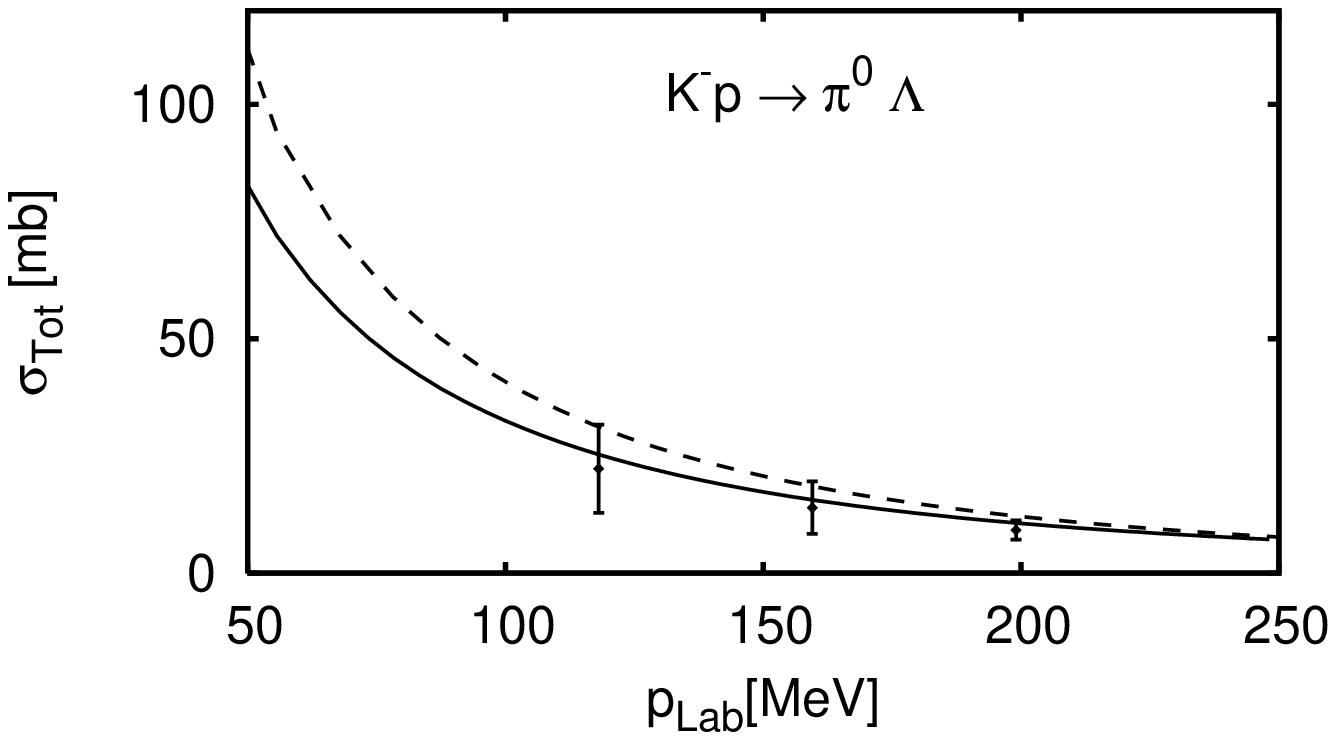}}
\caption{The total cross section of 
$K^-p \rightarrow K^-p$ (Left),
$K^-p \rightarrow \pi^+\Sigma^-$ (Center) and
$K^-p \rightarrow \pi^0\Lambda$ (Right) reactions.
The solid (dashed) curves shows the result using relativistic
(non-relativistic) model. Data are taken from Ref. \cite{crs1,crs2,crs3,crs4,crs5}.
}
\label{kp-cros}
\end{figure*}

The cut off $\Lambda$ is determined so as to reproduce the
scattering length of $\bar{K}N$ by Martin\cite{Mart},
which is summarized in Table \ref{hyo3}.
We have two models for the  non-relativistic and relativistic
kinematic energies. 
The relativistic form of the kinetic energy may be necessary 
due to for the $\pi\Sigma N$ channel
small pion mass.
The scattering length of $I=0$ is close to the value
$-1.70 + i0.68(fm)$ of Ref. \cite{Mart} but the real part
of our  $I=1$ scattering length is a little bit larger than $0.37 + i0.60(fm)$
of Ref. \cite{Mart}. They are consistent with the data
of the kaonic hydrogen atom\cite{Iwa,Itoh,Dear}. 
In $\bar{K}N-\pi\Sigma$ for $I=0$ channel, two models have
a resonance in $\bar{K}N$ physical and $\pi\Sigma$ unphysical sheet.
Both models give satisfactory description of the total
cross section of $K^-p$ reaction at low energy shown in Fig. \ref{kp-cros}.

\begin{table*}[htbp]
\begin{center}
\begin{tabular}{c||c|ccccc}
     &     &$\bar{K}N$(MeV)&$\pi\Sigma$(MeV)&$\pi\Lambda$(MeV)
     & Scattering Length(fm) & Resonance energy(MeV)\\ \hline \hline
Relativistic Model & I=0 & 1125 & 1300 &   & $-1.71+i0.56$ & $1413.6-i29.0$\\
      & I=1 & 1100 & 1100 & 1100 & $\ \ 0.66+i0.64$  &\\ \hline
Non-rela. Model  & I=0 & 960 & 900 &   & $-1.78+i0.59$ & $1414.3-i26.4$\\ 
        & I=1 & 850 & 950 & 900 & $\ \ 0.78+i0.66$  &\\ \hline
\end{tabular}
\caption{The relativistic and non-relativistic models of
the $\bar{K}N$ interaction.}
\label{hyo3}
\end{center} 
\end{table*}

\section{Results and Discussion}

We have searched resonance pole of the $\bar{K}NN-\pi\Sigma N$
coupled channel equation using a method described in section 2 and
the $\bar{K}N$ interaction explained in section 3. 
In addition, $NN$ interaction for $^1S_0$
channel and $\pi N$ interaction are included in the AGS-equation.

\begin{figure}
\resizebox{0.4\textwidth}{!}{%
\includegraphics{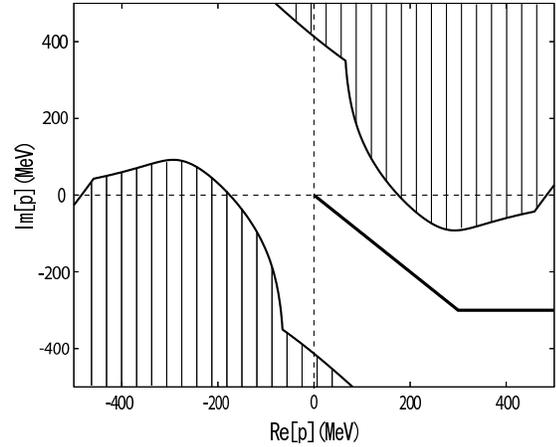}}
\caption{Deformed contour of the momentum integration.}
\label{Fig-ppath}
\end{figure}

In order to investigate the resonance position,
we have to analytically continue the equation into unphysical energy
sheet. For this purpose we deform the contour
of momentum integration so that we will not cross the
singularity of the $Z$ and $\tau$ of kernel.
The singularities are due to the $\pi\Sigma N, \bar{K}NN$ continuum in
the Green function and
$\bar{K}N$ two-body resonance corresponding to $\Lambda(1405)$,
and singularity of the potential for the complex momentum.
As an example the momentum contour is shown in 
solid curve in Fig. \ref{Fig-ppath} for 
$W=2m_N + m_K -70-i32MeV$.
We search resonance energy  below $\bar{K}NN$ 
and above $\pi\Sigma N$ threshold on the 
$\bar{K}NN$-physical and $\pi\Sigma N$-unphysical Riemann-sheet.
The shaded area is 'forbidden region' due to the singularity 
of $Z$ for kaon exchange mechanism.
We have studied all 'forbidden region' for $\pi, N$ and $K$
exchange mechanism and determined the integration contour.

\begin{figure}
\resizebox{0.45\textwidth}{!}{%
\includegraphics{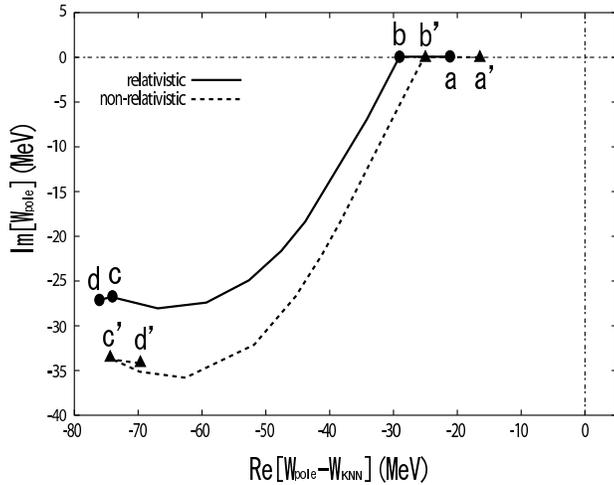}}
\caption{The pole trajectories of the $\bar KNN-\pi\Sigma N$ scattering
amplitude for the $J^{\pi}=0^-$ and $T=1/2$ state.
Two trajectories correspond to the relativistic model
(solid line and filled circles)
and non-relativistic one
(dashed line and filled triangles).}
\label{Fig-traject}
\end{figure}

At first, we take into account only $\bar KN-\bar KN$ interaction.
Therefore coupling with $\pi\Sigma N $ channel is switched off and
the contour of the  momentum integration is on the real axis.
In this case, we find bound state pole of the AGS-equation
below $\bar{K}NN$ threshold on physical sheet.
The results are shown in Fig. \ref{Fig-traject}
marked by $a$ and $a'$ for the 'relativistic' and 'non-relativistic' model.
Then including $NN$ interaction,  the binding energy is further
increased 
as 29.1MeV (25.2MeV)  at $b$ ($b'$) for 'relativistic'
('non-relativistic') model.
The $\bar{K}N$ interaction included in $\tau$ and $Z$
in this model is strong enough to bind  $\bar{K}NN$ system.

In the next step, we take into account the channel
coupling between $\bar KN$ state and $\pi\Sigma$ state
keeping kaon and nucleon exchange mechanism in $Z$.
We trace the trajectory of the resonance pole by artificially modifying
the strength of $\bar KN-\pi\Sigma$ transition potential 
from zero to the value of our model.
The solid curve and dashed curve in Fig. \ref{Fig-traject}
represent the pole trajectory 
corresponding to relativistic and non-relativistic model. 
The bound state pole moves into the $\bar KNN$ physical
and $\pi\Sigma N$ unphysical energy sheet
and reaches to $c$ ($c'$).
The width of the resonance is due to the decay of $\bar{K}NN$ bound state
to the $\pi\Sigma N$ and $\pi\Lambda N$ states
through the imaginary part of the  $\tau$.
Finally we include  the
$\pi$ exchange mechanism in $Z$ 
and $\pi -N$ two-body scattering terms in $\tau$,
which plays rather minor role in determining the pole position.
The final result of the  $\bar KNN-\pi\Sigma N$ resonance poles 
denoted by $d$ and $d'$ in Fig. \ref{Fig-traject}.
The the pole position of the three-body resonance
is $W = M - i \Gamma/2 = 2m_N+m_k - 76.1 - 27.1i$MeV 
($2m_N + m_K - 69.7 - 34.2i$)
for relativistic (non-relativistic) model.
Our resonance has deeper binding energy and similar width compared
with the prediction of Ref. \cite{Aka2}.
Recently Shevchenko, Gal and Mares\cite{she} studied $K^-pp$ system using
coupled channel Faddeev equation, which is quite similar approach
as our present study. They reported  $B \sim 55 - 70$MeV and
$\Gamma \sim  95 - 110$MeV. 
Their binding energy is similar to ours, while
our relativistic model gives smaller width.

In summary we have studied strange dibaryon state using the
$\bar{K}NN-\pi\Sigma N$ coupled channel Faddeev equation.
We found a resonance pole of the strange dibaryon
at $B \sim 76$MeV and $\Gamma \sim 54$MeV
in relativistic model. 
It is however noticed that the  $\bar{K}N$ interaction is 
not well determined experimentally and further investigation
to find a possible  the range of the resonance energy is necessary.
The full content of our work 
will be reported elsewhere\cite{ikeda}.

The authors are grateful to Prof. A. Matsuyama for very useful
discussion on the three-body resonance.
This work is supported
by a Grant-in-Aid for Scientific Research on Priority Areas(MEXT),Japan
with No. 18042003 and 
the 21st Century COE Program, 
``Towards a New Basic Science:Depth and Synthesis''.
 
%%%%%%%%%%%%%%%%%%%%%%%%%%%%%%%%%%%%%%%%%%%%%%%%%%%%%%%%%%%%%%%%%%%%%%%%%%%%%%%%%%


\begin{thebibliography}{99}

%
\bibitem{Gal1} J. Mares, E. Friedman, and A. Gal, Nucl. Phys. A \textbf{770}, (2006) 84.
\bibitem{Ram} L. Tolos, A. Ramos and E. Oset, Phys. Rev. C \textbf{ 74}, (2006) 015203.

\bibitem{Aka1}Y. Akaishi and T. Yamazaki, Phys. Rev. C \textbf{ 65}, (2002) 044005.
\bibitem{Aka2}T. Yamazaki and Y. Akaishi, Phys. Lett. B \textbf{ 535}, (2002) 70.
\bibitem{Dote}A. Dote et al., Phys. Rev. C \textbf{ 70}, (2004) 044313.
%
%
\bibitem{Agne}M. Agnello et al., Phys. Rev. Lett. \textbf{ 94}, (2005) 212303.
%
%
\bibitem{Oset2}E. Oset and H. Toki, Phys. Rev. C \textbf{ 74}, (2006) 015207.
\bibitem{Magas}V.K. Magas et al., Phys. Rev. C \textbf{ 74}, (2006) 025206.
%
\bibitem{book} I.R. Afnan and A.W. Thomas, \textit{Modern Three-Hadron
       Physics} (Springer, Berlin, 1977) Chap. 1
%
\bibitem{Glock}W. Gl${\rm \ddot{o}}$ckle, Phys. Rev. C \textbf{ 18}, (1978) 564. 
\bibitem{Bal}V. B. Belyaev and K. M${\rm \ddot{o}}$ller, Z. Phys. A
	\textbf{279}, (1976) 47.
\bibitem{Moll}K. M${\rm \ddot{o}}$ller, Czech. J. Phys. \textbf{ 32}, (1982) 291.
\bibitem{Matsu}A. Matsuyama and K. Yazaki, Nucl. Phys. A \textbf{ 534}, (1991) 620.
A. Matsuyama, Phys. Lett. B \textbf{408}, (1997)  25.
\bibitem{Pear}B.C. Pearce and I.R. Afnan, Phys. Rev, C \textbf{ 30}, (1984) 2022.
\bibitem{Afnan}I.R. Afnan and B.F. Gibson, Phys. Rev. C \textbf{ 47}, (1993) 1000.
%
%
\bibitem{Jido}D. Jido et al., Nucl. Phys. A \textbf{ 725}, (2003) 181.

\bibitem{Bor}B. Borasoy, R. Nissler and W. Weise, Eur. Phys. J. A \textbf{ 25}, (2005) 79.

\bibitem{Hama}T. Hamaie, M. Arima, and K. Masutani, Nucl. Phys. A \textbf{ 591}, (1995) 675.
%
%
\bibitem{ags}E. O. Alt, P. Grassberger and W. Sandhas, Nucl. Phys. B { 2}, (1967), 167.
%
%
\bibitem{Afnan2} I.R. Afnan and A.W. Thomas, Phys. Rev. C \textbf{ 10}, (1974) 109.
%
%
\bibitem{Mart}A.D. Martin, Nucl. Phys. B \textbf{ 197}, (1981) 33.
%
\bibitem{Iwa}M. Iwasaki et al., Phys. Rev. Lett. \textbf{ 78}, (1997) 3067.
\bibitem{Itoh} T. M. Itoh et al., Phys. Rev. C \textbf{58}, (1998) 2366.
\bibitem{Dear}G. Beer et al., Phys. Rev. Lett. \textbf{94}, (2005) 212302.
\bibitem{crs1}W. E. Humphrey and R. R. Ross, Phys. Rev. \textbf{127},
	(1962) 1305.
\bibitem{crs2}M. Sakitt et al., Phys. Rev. \textbf{139}, (1965) B719.
\bibitem{crs3}J. K. Kim, Phys. Rev. Lett. \textbf{14}, (1965) 29.
\bibitem{crs4}W. Kittel, G. Otter and I. Wacek, Phys. Lett. \textbf{21},
	(1966) 349.
\bibitem{crs5}D. Evans et al., J. Phys. G \textbf{9}, (1983) 885.
%
%
%
\bibitem{she}N. V. Shevchenko, A. Gal and J. Mares, nucl-th/0610022
\bibitem{ikeda} Y. Ikeda and T. Sato, in preparation.

%
\end{thebibliography}
\end{document}